\documentclass{article}

\PassOptionsToPackage{numbers, compress}{natbib}

\usepackage[preprint]{nips_2018}

\usepackage[utf8]{inputenc} 
\usepackage[T1]{fontenc}    
\usepackage{hyperref}       
\usepackage{url}            
\usepackage{booktabs}       
\usepackage{amsfonts}       
\usepackage{nicefrac}       
\usepackage{microtype}      
\usepackage[pdftex]{graphicx}

\title{Improved Speech Enhancement with the Wave-U-Net}

\author{
  Craig Macartney \\
  City, University of London\\
  \texttt{craig.macartney@city.ac.uk} \\
  \And
  Tillman Weyde \\
  City, University of London \\
  \texttt{t.e.weyde@city.ac.uk} 
}

\begin{document}

\maketitle

\begin{abstract}
 We study the use of the Wave-U-Net architecture for speech enhancement, a model introduced by Stoller et al for the separation of music vocals and accompaniment. 
 This end-to-end learning method for audio source separation 
 operates %
 directly in the time domain, permitting the integrated modelling of phase information 
 and being able to take large temporal contexts into account. 
 Our experiments show that the proposed method improves several metrics, namely PESQ, CSIG, CBAK, COVL and SSNR, 
 over 
 the 
 state-of-the-art with respect to the speech enhancement task on the Voice Bank corpus (VCTK) dataset. 
 We find that a reduced number of hidden layers is sufficient for speech enhancement in comparison to the original system designed for singing voice separation in music.  
 We see this initial result as an encouraging signal to further explore speech enhancement in the time-domain, both as an end in itself and as a pre-processing step to speech recognition systems. 
\end{abstract}

\section{Introduction}

Audio source separation refers to the problem of extracting one or more target sources while suppressing interfering sources and noise \citep{2018AudioEnhancement}. 
Two related tasks 
are those of speech enhancement and singing voice separation, both of which involve extracting the human voice as a target source. 
The former involves attempting to improve  speech intelligibility and quality when obscured by additive noise \citep{Loizou:2013:SET:2484638, Pascual2017, 2018AudioEnhancement}; whilst the latter's focus is on separating music vocals from accompaniment \citep{Stoller2018Wave-U-Net:Separation}.

Most audio source separation methods operate not directly in the time-domain, but with 
time-frequency representations as input and output (front-end).
Since 2017, the U-Net architecture on magnitude spectrograms has achieved new state of the art results in audio source separation for music \citep{Jansson-et-al-2017-Sining} and speech dereverbration \citep{Ori-et-al-2018-Speech-Dereverbration}. 
Also, neural network architectures operating in the time domain have recently been proposed for speech enhancement \citep{Pascual2017, RethageADenoising}.
These approaches have been combined in the Wave-U-Net
\citep{Stoller2018Wave-U-Net:Separation} and applied to singing voice separation. 
In this paper we apply the Wave-U-Net to speech enhancement and show that it produces results that are better than the current state of the art. 

The remainder of this paper is structured as follows. In section~\ref{sec:related}, we briefly review related work from the literature. 
In section~\ref{sec:wave-u-net}, we introduce briefly the Wave-U-Net architecture and its application to speech. 
Section \ref{sec:experiments} presents the experiments we conducted and their results including comparison to other methods. 
Section~\ref{sec:conclusions} concludes this article with a final summary and perspectives for future work.

\section{Related work}\label{sec:related}
Source separation of audio has seen great improvement in recent years through deep learning models  \citep{huang2015joint, nugraha2016multichannel}. 
These methods, as well as more traditional ones, mostly operate in the time-frequency domain, from deep recurrent architectures predicting soft masks, such as \citep{HuangSINGING-VOICENETWORKS}, to convolutional encoder-decoder architectures like that of \citep{Chandna2017MonoauralNetworks}.
Recently, the U-Net architecture on magnitude spectrograms has achieved new state of the art results in audio source separation for music \citep{Jansson-et-al-2017-Sining} and speech dereverbration \citep{Ori-et-al-2018-Speech-Dereverbration}. 

Also recently, models operating in the time domain have been developed.
The development of Wavenet \citep{Oord-et-al-2016-Wavenet} inspired other developments, including  \citep{Pascual2017, 
RethageADenoising}.
The SEGAN \citep{Pascual2017} architecture was developed for the purpose of speech enhancement and denoising. 
It employs a neural network in the time-domain with an encoder and decoder pathway that successively halves and doubles the resolution of feature maps in each layer, respectively, and features skip connections between encoder and decoder layers. 
It offers state-of-the-art results on the Voice Bank (VCTK) dataset \citep{Valentini-Botinhao2017Noisysound}.





The Wavenet for Speech Denoising \citep{RethageADenoising}, another architecture to operate directly in the time domain, takes its inspiration from \citep{Oord-et-al-2016-Wavenet}. It has a non-causal conditional input and a parallel output of samples for each prediction and is based on the repeated application of dilated convolutions with exponentially increasing dilation factors to factor in context information.





\section{Wave-U-Net for Speech Enhancement}\label{sec:wave-u-net}



\begin{figure}
    \centering
    \includegraphics[width=0.8\textwidth]{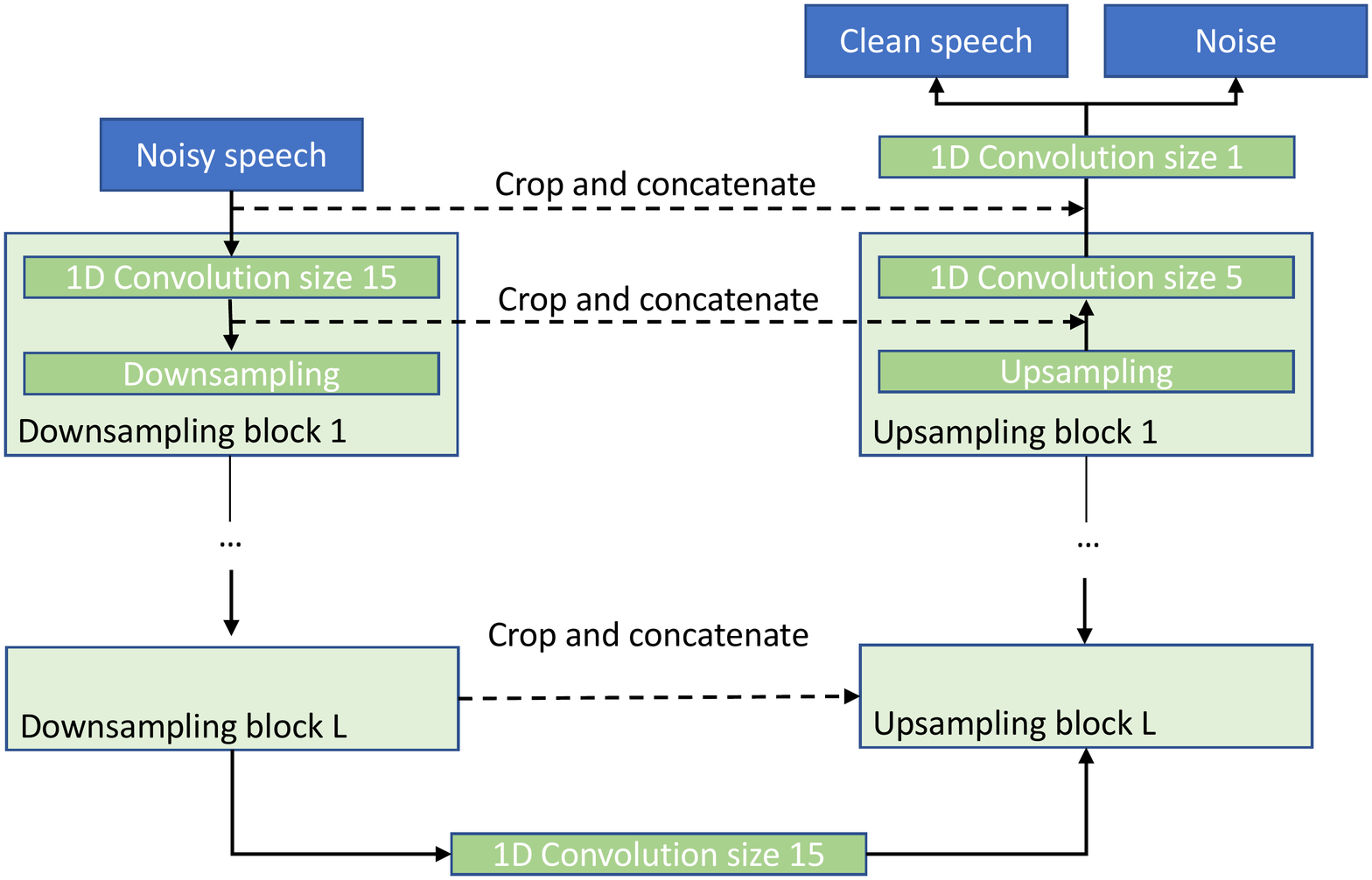}
    \caption{The Wave-U-Net architecture following \citep{Stoller2018Wave-U-Net:Separation}.}
    \label{fig:arch}
\end{figure}

The Wave-U-Net architecture of \citep{Stoller2018Wave-U-Net:Separation} combines elements of both of the abovementioned architectures with the U-Net. 
The overall architecture is a one-dimensional U-Net with down and upsampling blocks.

As per the spectrogram-based U-Net architectures (e.g. \citep{Jansson-et-al-2017-Sining}), the  Wave-U-Net uses a series of downsampling and upsampling blocks to make its predictions, whilst at each level of the network, the time resolution is halved. 


In applying the Wave-U-Net architecture to the application of speech enhancement, our objective is to separate a mixture waveform $m \in [-1, 1]^{L \times C}$ into K source waveforms
$S^1, . . .,S^K$
with $S^k \in [-1, 1]^{L \times C}$ for all $k \in {1, . . . , K}, C$ 
as the number of audio channels and $L$ as the numbers of audio samples. 
In our case of monaural speech enhancement we have $K = 2$ and $C = 1$.

In doing so, for K sources to be estimated, a 1D convolution, zero-padded before convolving, of filter size 1 with K · C filters is applied to convert the stack of features at each audio sample into a source prediction for each sample. This is followed by a tanh nonlinearity to obtain a source signal estimate with values in the interval $(-1, 1)$. All convolutions except the final ones are followed by LeakyReLU non-linearities.

\section{Experiments} \label{sec:experiments}

\subsection{Datasets}

We use the same VCTK dataset \citep{Valentini-Botinhao2017Noisysound} as the SEGAN \cite{Pascual2017}, which is available publicly, encouraging comparisons with future speech enhancement methods.

The dataset includes clean and noisy audio data at 48kHz sampling frequency.
However, like the SEGAN, we downsample to 16kHz for training and testing.
The clean data are recordings of sentences, sourced from various text passages, uttered by 30 English-speakers, male and female, with various accents – 28 intended for training and 2 reserved for testing \citep{Valentini-Botinhao2016SpeechNetworks}. 
The noisy data were generated by mixing the clean data with various noise datasets, as per the instructions provided in \citep{Pascual2017, Valentini-Botinhao2017Noisysound, Valentini-Botinhao+2016}. 

With respect to the training set, 40 different noise conditions are considered \citep{Pascual2017, Valentini-Botinhao2016SpeechNetworks}. 
These are composed of 10 types of noise (2 of which are artificially-generated\footnote{available here: http://data.cstr.ed.ac.uk/cvbotinh/SE/data/ \citep{Valentini-Botinhao2017Noisysound}} and 8 sourced from the DEMAND database \citep{Thiemann2013TheRecord-ings}, each mixed with clean speech at one of 4 signal-to-noise ratios (SNR) (15, 10, 5, and 0 dB). 
In total, this yields $11,572$ training samples, with approximately 10 different sentences in each condition per training speaker.

Testing conditions are mismatched from those of the training. The speakers, noise types and SNRs are all different. The separate test set with 2 speakers, unseen during training, consists of a total of 20 different noise conditions: 5 types of noise sourced from the DEMAND database at one of 4 SNRs each (17.5, 12.5, 7.5, and 2.5 dB) \citep{Valentini-Botinhao2017Noisysound, Valentini-Botinhao+2016}. This yields 824 test items, with approximately 20 different sentences in each condition per test speaker \citep{Valentini-Botinhao2017Noisysound, Valentini-Botinhao+2016}.

\subsection{Experimental setup}

As per \citep{Stoller2018Wave-U-Net:Separation}, our baseline model trains on randomly-sampled audio excerpts, using the ADAM optimization algorithm, a learning rate of 0.0001, decay rates $\beta1$ = 0.9 and $\beta2$ = 0.999 and a batch size of 16. 
We specify  
an initial network layer size of 12, like in \citep{Stoller2018Wave-U-Net:Separation}, although this is varied across experiments, as described in the Results section below. 16 extra filters per layer are also specified, with downsampling block filters of size 15 and upsampling block filters of size 5 like in \citep{Stoller2018Wave-U-Net:Separation}. 
We train for 2,000 iterations with mean squared error (MSE) over all source output samples in a batch as loss and apply early stopping if there is no improvement on the validation set for 20 epochs.  
We use a fixed validation set of 10 randomly selected tracks.
Then, the best model is fine-tuned  with the batch size doubled and the learning rate lowered to 0.00001, again until 20 epochs have passed without improved  validation loss.

\subsection{Results}\label{subsec:results}

\begin{table}
  \caption{Objective evaluation - comparing the mean results of the untreated noisy signal, the Wiener-, SEGAN- and Wave-U-Net-enhanced signals. Higher scores are better for all metrics.}
  \label{sample-table}
  \centering
  \begin{tabular}{lllll}
    \toprule
    \textbf{Metric}     &\textbf{Noisy}     &\textbf{Wiener}     &\textbf{SEGAN}     &\textbf{Wave-U-Net} \\
    \midrule
    PESQ & \,1.97 & \, \, 2.22 & \, \, 2.16 & \textbf{\, \, \, \, \, 2.40}    \\
    CSIG & \,3.35 & \, \, 3.23 & \, \, 3.48 & \textbf{\, \, \, \, \, 3.52}     \\
    CBAK & \,2.44 & \, \, 2.68 & \, \, 2.94 & \textbf{\, \, \, \, \, 3.24}     \\
    COVL & \,2.63 & \, \, 2.67 & \, \, 2.80 & \textbf{\, \, \, \, \, 2.96}     \\
    SSNR & \,1.68 & \, \, 5.07 & \, \, 7.73 & \textbf{\, \, \, \, \, 9.97}     \\
    
    \bottomrule
  \end{tabular}
\end{table}

\begin{table}
  \caption{Objective evaluation - mean results, comparing variations of the Wave-U-Net model with different numbers of layers, without fine-tuning applied.}
  \label{sample-table}
  \centering
  \begin{tabular}{llllll}
    \toprule
    \textbf{Metric}     &\textbf{12-layer}     &\textbf{11-layer}     &\textbf{10-layer}     &\textbf{9-layer}     &\textbf{8-layer} \\
    \midrule
    
    PESQ & \, \, 2.40 & \, \, 2.38 & \textbf{\, \, 2.41} & \textbf{\, \, 2.41} & \, \, 2.39    \\
    CSIG & \, \, 3.49 & \, \, 3.47 & \, \, 3.43 & \textbf{\, \, 3.54} & \, \, 3.51      \\
    CBAK & \, \, 3.23 & \, \, 3.22 & \textbf{\, \, 3.24} & \, \, 3.23 & \, \, 3.18      \\
    COVL & \, \, 2.95 & \, \, 2.92 & \, \, 2.92 & \textbf{\, \, 2.97} & \, \, 2.95    \\
    SSNR & \, \, 9.79 & \, \, 9.95 & \textbf{\, \, 9.98} & \, \, 9.87 & \, \, 9.30     \\
    
    \bottomrule
  \end{tabular}
\end{table}

To evaluate and compare the quality of the enhanced speech yielded by the Wave-U-Net, we mirror the objective measures provided in \citep{Pascual2017}. 
Each measurement compares the enhanced signal with the clean reference of each of the 824 test set items. 
They have been calculated using the implementation provided in \citep{Loizou:2013:SET:2484638}\footnote{available here: https://ecs.utdallas.edu/loizou/speech/software.htm}. 
The first metric is that of the Perceptual Evaluation of Speech Quality (PESQ) - more specifically the wide-band version recommended in ITU-T P.862.2 (from –0.5 to 4.5) \citep{Loizou:2013:SET:2484638, Pascual2017}. Secondly, composite measures of metrics that aim to computationally approximate the Mean Opinion Score (MOS) that would be produced from human perceptual trials are computed \citep{RethageADenoising}. These are: CSIG, a prediction of the signal
distortion attending only to the speech signal \citep{Hu2008EvaluationEnhancement} (from 1 to 5); CBAK, a prediction of the intrusiveness of background noise \citep{Hu2008EvaluationEnhancement} (from 1 to 5); and COVL, a prediction of the overall effect \citep{Hu2008EvaluationEnhancement} (from 1
to 5). Last is the Segmental Signal-to-Noise Ratio (SSNR) \citep{Quackenbush1988ObjectiveQuality} (from 0 to $\infty$).

Table 1 shows the results of these metrics for comparison across different speech enhancement architectures versus the overall best-performing Wave-U-Net model, that of the 10-layer with fine-tuning applied. 
As a comparative reference, it also shows the results of these metrics when applied: directly to the noisy signals; to signals filtered using the Wiener method, based on a priori SNR estimation; and to the SEGAN-enhanced signal, as provided in \citep{Pascual2017}. 
The results indicate that the Wave-U-Net is the most effective model for speech enhancement.

Table 2 shows the performance differences between different variations of the Wave-U-Net, 
with different numbers of layers. 
No fine-tuning was performed to obtain the results shown here, which explains the difference between the 10-layer Wave-U-Net in Tables 1 and 2. 
The results suggest that fine-tuning does not make a meaningful difference, except on the CSIG measurement, and that performance reaches a peak around the 10- and 9-layer models, smaller than the best performing equivalent model for music vocals source separation in \citep{Stoller2018Wave-U-Net:Separation}.
This is likely due to the size of the receptive field, where for speech the optimal size is probably smaller than for music.




\section{Conclusions}\label{sec:conclusions}

\subsection{Summary}
The Wave-U-Net combines the advantages of several of the most recent successful architectures for music and speech source separation and our results show that it is particularly effective at speech enhancement.  
The results improve over the state of the art by a good margin even without significant adaptation or parameter tuning. 
This indicates that there is great potential for this approach in speech enhancement.

\subsection{Future work}
In comparison to the SEGAN architecture, it is possible that the advantage stems from the upsampling that avoids aliasing, which should be further investigated. 
The results indicate that there is room for increasing effectiveness and efficiency by further adapting the model size and other parameters, e.g. filter sizes, to the task and expanding to multi-channel audio and multi-source-separation.

\small
\bibliographystyle{plainnat}
\bibliography{refs2} 






\end{document}